\documentclass[11pt,hyper,twocolumn]{article}
\usepackage{amsmath,epsfig}
\usepackage{amssymb,amsfonts}
\usepackage{abstract}
\usepackage{epstopdf}
\usepackage{graphicx}
\usepackage{multirow,cite}
\usepackage{subfigure}
\usepackage{color}
\usepackage{datetime}

\newcommand{\bi}{\begin{itemize}}
\newcommand{\ei}{\end{itemize}}
\newcommand{\non}{\nonumber}
\def\p{\partial}
\def\a{\alpha}

\def\g{\gamma}
\def\l{\lambda}

\def\s{\sigma}

\def\O{\mathcal{O}}

\def\D{\Delta}

\def\sl2r{SL(2,\mathbb{R})}

\def\r{\rightarrow}
\def\half{{\frac12}}

\newcommand{\bea}{\begin{eqnarray}}
\newcommand{\eea}{\end{eqnarray}}
\newcommand{\be}{\begin{equation}}
\newcommand{\ee}{\end{equation}}

\title{\bf Bulk fields from the boundary OPE \vspace{4mm}}

\author{
Monica Guica\\
\\\vspace{2mm}
\emph{Institut de Physique Th\'eorique, CEA Saclay},
\emph{91191 Gif-sur-Yvette, France}\\\vspace{2mm}
\emph{ Department of Physics and Astronomy, Uppsala University,   SE-751 08 Uppsala, Sweden}\\
 \emph{ Nordita, Stockholm University and KTH Royal Institute of Technology,}\\
\emph{Roslagstullsbacken 23, SE-106 91 Stockholm, Sweden}
}

\date{}
\begin{document}

\twocolumn[
  \begin{@twocolumnfalse}
    \maketitle
    \begin{abstract}
      Previous work has established an equality between the geodesic integral of a free bulk field in AdS and the contribution of the conformal descendants of its dual CFT primary operator to the OPE of two other operators inserted at the endpoints of the geodesic. Working in the context of the AdS$_3$/CFT$_2$ correspondence, we  extend this relation to include the $1/N$ corrections to the bulk field obtained by dressing it with i) a $U(1)$ current and ii) the CFT stress tensor. In the former case, we argue that the contribution of the Ka\v{c}-Moody descendants to the respective boundary OPE equals the geodesic integral of a particular $U(1)$-dressed bulk field, which is framed to the boundary via a split Wilson line. In the latter case, we  compute the gravitational $1/N$ corrections to the bulk field in various gauges, and then write a  CFT expression for a putative bulk field whose geodesic integral captures the contribution of Virasoro descendants to the  OPE of interest. We comment on the 
       bulk interpretation of this expression.

    \end{abstract}
    
\vspace{0.7cm}
  \end{@twocolumnfalse}
  ]

\maketitle

\noindent It has been recently observed \cite{Czech:2016xec,deBoer:2016pqk,daCunha:2016crm} that the contribution of the conformal family of a primary operator $\O$ in a CFT$_d$ to the OPE of two other primaries $A,B$ \cite{Ferrara:1971vh} can be represented as  the integral of a  free  bulk field in AdS$_{d+1}$ constructed from $\O$  \cite{Bena:1999jv,Hamilton:2006az,Hamilton:2006fh} 
\be 
\Phi^{(0)}(y) = \int d^dx\,  K_\D (y|x) \O (x)
\ee
over the geodesic $\g_{AB}$ that joins the boundary insertions of $A,B$
\be
A(x) B(0) \sim \frac{
C_{AB\O}}{|x|^{\D_A +\D_B}} \int_{\g_{AB}} d\l \, e^{-\l \D_{AB}}\, \Phi^{(0)} (y(\l))  \label{fgg}
\ee
The above is a purely kinematic relation, which simply follows from conformal symmetry and is thus valid in any CFT. In a large $N$ CFT, $\Phi^{(0)}$ is the leading approximation at large $N$ to the physical bulk field, which in general involves an
 infinite series of multitrace corrections \cite{Kabat:2011rz,Kabat:2016zzr}
\bea
\Phi_{HKLL} & = & \int K_\D \O + \frac{1}{N} \sum_n c_n \int K_{\D_n} (AB)_n + \non \\ &&\hspace{-1.5cm} +\frac{1}{N^2}\sum_{m,n} d_{m,n}\int K_{\D_{m,n}} (C(AB)_n)_m + \ldots  \label{hkll}
\eea
whose coefficients are fixed by the OPE coefficients of $\O$. Given this fact, it is interesting to ask whether a relation of the form \eqref{fgg} could hold between the full $A B$ OPE and the all-orders bulk field \eqref{hkll}. 

In this note, we would like to verify this relation for a particular subset of operators that must appear in the $AB$ OPE together with $\O$, namely multitrace operators constructed from one $\O$ and an arbitrary number of insertions of either a $U(1)$ current, $j$, 
 or of the CFT stress tensor, $T$. We specialize to the context of two-dimensional CFTs, where the couplings of these operators are universal and  entirely determined by the associated Ka\v{c}-Moody/Virasoro symmetry\footnote{That such a relation should exist is suggested by the fact  that the Virasoro  heavy-light conformal  block, which
sums up the above multitrace contributions,  is the geodesic
integral of the corresponding bulk field  \cite{Hijano:2015qja}.}.  The dual bulk theory is described by a scalar field coupled to $U(1)$ Chern-Simons theory or to pure three-dimensional Einstein gravity, in a limit where the backreaction  of the scalar on the bulk gauge field/the metric is entirely neglected. As a consequence, the bulk gauge field/metric is pure (large) gauge and the  dressing  of the bulk scalar by the current/stress tensor can be computed to all orders.

\vskip1.5mm

\noindent \emph{Note added in v2:} After this article appeared on arXiv, it was brought to our attention that  \cite{lamprou} were also pursuing similar ideas. 

\bigskip

Let us first concentrate on  the case of a charged bulk scalar in AdS$_3$ coupled to a $U(1)$ Chern-Simons gauge field, which is rendered gauge-invariant by attaching to it  a Wilson line running from the bulk point $y$ to a boundary point $x_0$. The expression for the gauge-invariant bulk field operator, to linear order in $\O$ and all orders in the current, has been worked out in  \cite{wdan} 
\be
\widehat \Phi_J (y) = \int d^2 x  \, K_\D(y|x) \,
e^{-i q \int_{x_0}^x j(x') dx'} \O(x) 
\ee
where $q$ is the $U(1)$ charge of $\O$ and $\D$ is its conformal dimension.  Note that $\widehat \Phi_J (y)$ still satisfies the free wave equation. 

We would like to know whether the OPE of two boundary operators, A, B in  which one includes both the (global) conformal and the Ka\v{c}-Moody descendants of the operator $\O$,   equals  the geodesic integral of this type of gauge-invariant bulk field for some choice of the Wilson line used to frame it to the boundary\footnote{Note that in the approximation in which we are working, the shape of the Wilson line is irrelevant, and the framing is entirely specified by its boundary endpoint(s). }. For this, it is useful to bosonize the current to a chiral boson operator, $\chi(z)$, via $\p \chi (z) = j(z)$. The OPE of $\chi$ with various operators can be derived from the known $j$ OPEs. As explained at length in \cite{wdan}, while generically $\chi$ does not exist as an operator in the holographic dual CFT (since its zero mode is unphysical),  it is nevertheless a useful computational tool. In terms of $\chi$, the expression for the gauge-invariant bulk scalar reads
\be
\widehat \Phi_J (y) = \int d^2 x  \, K_\D(y|x) \,e^{iq (\chi(x_0)-\chi(x))} \O (x) \label{bfcs}
\ee


\noindent The way we will be checking the relation between the all-orders bulk field \eqref{bfcs} and the boundary OPE is by computing correlators of the form $\langle j(w_1) \ldots j(w_n) \, A(x_1)\, B(x_2)\,  \O^\dag (x_3) \rangle$ with $x_i =(z_i,\bar z_i)$. In a CFT$_2$, all these correlators can be obtained recursively from the $\langle A(x_1) B (x_2)  \O^\dag (x_3)\rangle$ three-point function, which is non-zero provided that $q_A + q_B = q$. For example, the  correlator with one insertion of $j$ is
\vspace{-1mm}
\bea
&& \hspace{-8mm}\langle j(w) \, A(x_1) B(x_2)  \,\O^\dag (x_3 ) \rangle = \left( \frac{q_A}{w-z_1} +\frac{q_B}{w-z_2} \right. \non  \\
&& \hspace{-5mm} \left.  -  \frac{q}{w-z_3} \right) \langle  A(x_1) B(x_2) \,\O^\dag (x_3 ) \rangle  \label{j4pf} 
\eea
which should equal the $\O(q)$ term of
\vspace{-1mm}
\bea
&& \hspace{-9mm} \int_{\g_{AB}} d\l\,  e^{-\l \D_{AB}} \int d^2 x' K_\D (y(\l)|x')  \times  \non \\ && \times \langle \O(x')\, e^{i q (\chi(x_0)-\chi(x'))} j(w) \, \O^\dag (x_3 )\rangle
\eea
where $q$ is taken to be $\O(1/N)$. The OPE of $j(w)$ with $ \O^\dag (x_3 )$ reproduces the last term in the correlator \eqref{j4pf}, whereas the OPE of $j (w)$ with $\O(x')$ cancels against the contribution of $\chi(x')$. Thus, the first two terms in \eqref{j4pf} must come from the OPE of $j(w)$ with $\chi (x_0)$.  If $q_B =0$, then this simply tells us that $x_0 = x_1$, i.e. the AB OPE equals the integral of the gauge-invariant bulk field framed by a Wilson line ending on the insertion of the operator $A$ on the boundary. However, if both $q_A, q_B \neq 0$, then the  bulk field should be connected to the boundary via a \emph{split} (or forked) Wilson line, which starts at $y$, splits in two, and ends on the boundary at $x_{1,2}$. We know such a Wilson line should exist, since  it contributes to the bulk three-point function of the operators with charges $q_A, q_B$  and $-q$. 

The expression for the gauge-invariant bulk operator framed this way is\footnote{Note that this dressing can  be entirely rewritten in terms of the current, using $q=q_A+q_B$, so the unphysical zero mode of $\chi$ cancels away.}
\bea
 \widehat \Phi^{AB}_J (y) & = & \int d^2 x  \, K_\D(y|x) \, \O (x) \times \non\\
 && \hspace{-15mm} \times \exp \left[ i (q_A \chi(x_1) + q_B \chi(x_2)-q \chi(x))\right] \label{bfj}
\eea
and it has, by construction, all the correct correlators with arbitrary powers of $j$ (since they are entirely determind by the singularity structure). Its dressing by $\chi$ can  also be understood in the shadow operator formalism \cite{Ferrara:1972uq} (see also \cite{Czech:2016xec,deBoer:2016pqk}), in  which the AB OPE is written as
\be
A(x_1) B(x_2) \sim \int_{\diamond} d^2 x  \, \langle A(x_1) B(x_2) \tilde \O(x)\rangle \, \O(x) \label{shadow}
\ee
where $\tilde \O$ is the shadow operator associated to $\O$. The split Wilson line in \eqref{bfj} is nothing but the  dressing by the current (or, equivalently, by $\chi$) of the  $\langle A B \tilde{O} \rangle$ correlator.


\bigskip

 Let us now turn to the gravitational case. We can find the all-orders  ``semiclassical'' gravitational dressing of a probe bulk scalar by considering its  propagation in a metric that solves the vacuum Einstein equations in AdS$_3$. The most general such metric
  is obtained by acting with a diffeomorphism $\xi^\mu$ on Poincar\'e  AdS$_3$
 
\be
ds^2 = \frac{dz^2 + dx^+ dx^-}{z^2} \label{poinc}
\ee 
Asymptotically, the diffeomorphism should approach
\bea
\xi & = & [f_+(x^+) +\O(z^2 )] \p_+ + [ f_-(x^-) + \O(z^2)] \p_- \non \\ &+&   \frac{z}{2} [\p_+ f_+(x^+) + \p_-  f_-(x^-)+ \O(z^2)] \p_z \label{xiasy}
\eea
If we require $\xi$ to preserve radial gauge, $g_{\mu z} =0$, then its form is entirely determined
\vspace{-2mm}
\bea
\xi_{rad} \! & = & \! [f_+(x^+) -\frac{z^2}{2}  f_-''(x^-)] \p_+ + [ f_-(x^-)  \\ && \hspace{-1.2cm} -\, \frac{z^2}{2} f_+''(x^+)] \p_- + \frac{z}{2} [\p_+ f_+(x^+) + \p_-  f_-(x^-)] \p_z\non 
\eea
where we are working at linearized level. In this approximation, $f_\pm$
are related to the expectation value of the stress tensor via $ \langle T_{\pm\pm} \rangle = \frac{c}{12} f_\pm^{(3)}$. 

In principle, we can find the bulk field by working perturbatively in $\xi^\mu$. At linear order, we have
\bea
(\Box-m^2) \, \Phi^{(1)} & = & 2 \nabla^\mu \left(\xi^\nu \nabla_\mu \nabla_\nu \Phi^{(0)}\right) -\non \\
&  & \hspace{-29mm} - 2 \,\xi^\nu \nabla_\nu \Box \, \Phi^{(0)} + \left(\Box + \frac{2}{\ell^2}\right) \xi^\nu \nabla_\nu \Phi^{(0)}
\eea
where $\ell$ is the AdS$_3$ radius and all derivatives are  with respect to the Poincar\'e AdS background. Inverting the propagator  as in \cite{Heemskerk:2012mn}, one finds
\bea
\Phi^{(1)}(y) &= & - \int d^2 x \, K_\D(y|x) \left( \xi^\a (x) \,\p_\a \O (x)\, + \right.\non \\&& \hspace{-15mm} \left.  + \,\,\frac{\D}{2}\, \p_\a \xi^\a (x) \O (x) \right) + \xi^\nu (y) \, \p_\nu \Phi^{(0)} (y) \label{fof}
\eea
where $\a = \pm$ runs over the boundary coordinates.

To find the linearized contribution to the gauge-invariant bulk field, one needs to add to \eqref{fof} the  contribution from the ``gravitational Wilson line'' used to render the operator gauge-invariant. The usual way  to specify a gauge-invariant bulk operator in gravity is as the bulk field at some fixed affine distance along a geodesic shot from the boundary \cite{Donnelly:2015hta}; more concretely, it is  $\widehat \Phi(y^\mu-\chi^\mu)$, where $\chi^\mu$ is an \emph{asymptotically trivial} diffeomorphism used to set the metric components along the said geodesic to zero. In general,  $\chi^\mu$ is solved for in terms of the metric; however, in our case it is simpler to write it in terms of the diffeomorphism $\xi^\mu$ used to generate the metric from pure AdS$_3$.  For example, to obtain the bulk field in radial gauge, one should take $\chi^\mu = \xi^\mu - \xi^\mu_{rad}$. In the appendix, we also work out the relevant $\chi^\mu$  for a bulk field framed along  certain ``circular'' geodesics.  

The above implies that the linearized correction to the  bulk field in radial gauge is given by \eqref{fof} with $\xi^\mu$ replaced by $\xi^\mu_{rad}$. Concentrating on the holomorphic contribution due to $f_+(x^+)$, the expression can be  massaged into
\bea
\widehat \Phi_{rad}^{(1)} (y) & = & \left(h-1\right) \int d^2 x' \s_0^{\D-3} \left(- \s_0 \,\Psi^{(2)}(x'|x) + \right.  \non\\
&+& \left. \frac{\D x^-}{z} \Psi^{(3)}(x'|x)  \right) \O(x') \label{phihrad}
\eea
where $\D = 2h$ and we used the explicit expression for the bulk-to-boundary propagator \cite{Hamilton:2006fh}
\be
K_\D (z,x|x') = \left(\frac{z^2 + \D x^+ \D x^-}{2z} \right)^{\D-2}\! \equiv \s_0^{\D-2}
\ee
with $\D x^\a = x'^\a-x^\a $. We also defined

\bea
\Psi^{(2)}(x'|x) & = & f'(x'^+) -f'(x^+) - f''(x^+) \D x^+  \non \\
\non \\
\Psi^{(3)}(x'|x) & = & f(x'^+) -f(x^+) - f'(x^+) \D x^+ - \non \\
&-&  f''(x^+)\, \frac{(\D x^+)^2 }{2} 
\eea
It is not hard to see that, at this order, 
\be
\Psi^{(2)}(x'|x) = \frac{12}{c} \int_{x^+}^{x'^+} dx''^+ \int_{x^+}^{x''^+} dx'''^+ T_{++} (x'''^+) \non
\ee
and 
\be
\Psi^{(3)}(x'|x) = \frac{12}{c} \int_{x^+}^{x'^+} \int_{x^+}^{x''^+}\int_{x^+}^{x'''^+} T_{++} \label{psifromt}
\ee
%
The linearized expression for the bulk field can then be written as
\bea
\widehat \Phi_{rad}^{(1)} (y) &=& - \int d^2 x' K_\D(y|x') \left( h \, \Psi^{(2)}(x'|x)  \O(x')  \right.\non \\
&+&  \left.  \Psi^{(3)}(x'|x) \p_{+'} \O(x') \right)
\eea
and it can be easily checked that it satisfies the linearized equation of motion
\be
(\Box-m^2) \, \widehat\Phi_{rad}^{(1)} = - \frac{24 z^4}{c}\, T_{++} \, \p_-^2 \Phi^{(0)}
\ee
Note that if we are turning on a purely holomorphic background ($f_-(x^-) =0$), then the expression for both the metric and its inverse truncates \cite{Skenderis:1999nb}, and the above equation of motion is valid to all orders in $T_{++}$.  In this case, it is not hard to guess the all-orders expression\footnote{
This is for a bulk field in radial gauge. In the appendix, we work out  the bulk field framed along a set of ``circular'' geodesics, and
 find an expression identical to \eqref{bft}, just with $x$ replaced by the boundary endpoint of the new geodesic.}\footnote{ Note   that    \eqref{bft} is not invariant under  special conformal transformations, as expected of a non-local operator in the bulk \cite{Kabat:2013wga}.}
\be
\widehat \Phi_{rad}
 (y) = \!\int \!\! d^2 x'   K_\D(y|x') \, e^{-h \Psi^{(2)}(x'|x)  -  \Psi^{(3)}(x'|x) \,\p_{+'}  }  \O(x') \label{bft}
\ee
with $\Psi^{(2),(3)}$ given by \eqref{psifromt}, which satisfies the equation of motion with a purey holomorphic source exactly. It is likely that the full expression for the gauge-invariant bulk field consists of the above holomorphic and a similar antiholomorphic exponentiated factor; however, it is harder to check that it satisfies the equation of motion, since the latter includes an infinite series in $ T_{++} T_{--}/c^2$.

\bigskip
Coming back to our original question, we are looking for a dressed  bulk field that, when integrated, has the same OPE with the boundary stress tensor as the AB OPE. If we are to follow the same strategy that we used in the $U(1)$ case, we should first  find a  dressing that makes the OPE of the bulk field with the CFT stress tensor be that of a local primary operator at some point $x_0$ on the boundary, irrespectively of its bulk position. In principle, we could first compute the OPE of \eqref{bft} with the stress tensor and understand how to make it behave as a local operator at $x_0$; however, in practice it is simpler to guess a CFT expression for a  bulk field   that  has the desired behaviour. 

For this, it is useful to introduce the Liouville field $\varphi$, related to the CFT stress tensor via \cite{Turiaci:2016cvo}
\be
T= Q \, \varphi'' - \varphi'^2  \;, \;\;\;\;\; Q = \sqrt{\frac{c}{6}} >>1
\ee
In terms of the holomorphic conformal transformation $\xi(x^+)$ that generates the background under study, the  stress tensor  and $\varphi$ are given by 
\be
T= \frac{c}{12} \left( \frac{3}{2} \frac{\xi''^2}{\xi'^2} - \frac{\xi'''}{\xi'} \right) \;, \;\;\;\;\;\; \varphi = \frac{Q}{2} \ln \xi' 
\ee
where we can recognize $T$ as a purely  Schwarzian derivative term. Given that  a Liouville vertex operator $e^{\a \varphi}$ behaves at large $Q$ as a primary of dimension $h = \frac{\a Q}{2}$,   the putative bulk field
\be
\widehat \Phi_T (y) =\int d^2 x \, K_\D(y|x)\,  e^{\frac{2h}{Q} \left( \varphi (x_0)-  \varphi(x)\right)}  \O(x) \label{bftl}
\ee
does have the desired OPE with the stress tensor, since the contribution of the integrated operator $\O(x)$ is canceled,  at leading order in $1/N$, by that of  $\exp[ - \frac{2h}{Q} \varphi(x)]$, which effectively contributes a ``negative dimension''. This behaviour is exactly analogous to that of the  charged bulk field \eqref{bfcs}. 

It is then not hard to write down a CFT expression for a bulk field whose OPE with the stress tensor is the same, at leading order in $1/N$, as that of an operator A inserted at $x_1$ and an operator B at $x_2$
\bea
\widehat \Phi^{AB}_T (x) &=& \int d^2 x \, K_\D(y|x)\, \O(x) \times \label{bftope} \\ && \hspace{-15mm} 
\times  \, \exp \left[\frac{2}{Q}(h_A \varphi(x_A) + h_B \varphi (x_B)- h \varphi(x)) \right] \non
\eea
This dressing is  natural also from the point of view of the shadow operator formalism, since $e^{2 \a/Q \varphi} = \xi'^\a$, so its role is to transform the integrated correlator in \eqref{shadow} from $\xi(x)$ to $x$  coordinates. It is then clear that the OPE relation we are looking for holds if the dressing of the integrated bulk field is given by \eqref{bftope}. 

While \eqref{bftl}, \eqref{bftope} represent scalar fields in the bulk in the sense that they satisfy the wave equation\footnote{Note that, unlike \eqref{bft}, they satisfy the \emph{free} wave equation in AdS, so now $x$ should be thought as the uniformizing coordinate in which $\langle T \rangle =0$, whereas in \eqref{bft} it simply represented the CFT coordinate.   } and have the correct $c\r \infty$ limit, we still need to specify which precise framing of the bulk field to the boundary yields these expressions. By analogy with the $U(1)$ case, we would like to interpret them as the bulk field attached to the boundary via a ``radial''  and respectively a ``split'' Wilson line. However, note that the ``radial Wilson line'' that frames \eqref{bftl} is different from the dressing of the  bulk field in radial gauge \eqref{bft}, even though at linear level they both involve  the double integral of the stress tensor. 
It seems reasonable to  attribute this discrepancy to the two distinct ways to define the gravitational dressing in $3d$: by labeling bulk points according to their affine distance to the boundary along geodesics, which is the method we used to derive \eqref{bft}, and by framing the bulk field with $SL(2,\mathbb{R})$ Wilson lines in the  Chern-Simons formulation of $3d$ gravity, which is our conjecture for what \eqref{bftl} represents. This conjecture is supported by the similarity between the bulk field dressing in \eqref{bftl} and the matrix elements of $SL(2,\mathbb{R})$ Wilson lines recently worked out in \cite{Fitzpatrick:2016mtp}. 
%
  By the same token, \eqref{bftope} would correspond to a split Wilson line dressing in the Chern-Simons formulation.



\bigskip

This concludes our argument for an equality between the contribution to the  boundary OPE of an operator $\O$ and all of its Ka\v{c}-Moody/Virasoro descendants and the geodesic integral of an appropriately-dressed bulk field operator. Since our proof ultimately relies on the fact that both the Ka\v{c}-Moody and the Virasoro descendants can be generated via a holomorphic gauge/coordinate transformation,
 it would be worthwhile to have a more explicit derivation of this relation, e.g. by summing the global multitrace primaries that make up the Virasoro block, along the lines of \cite{Fitzpatrick:2015foa}. 

Our work indicates that there are two physically different ways to define gauge-invariant operators in  $3d$ gravity, and it would be interesting to explore them further. First, one could check whether \eqref{bftl} corresponds indeed to the bulk field framed to the boundary via  an $SL(2,\mathbb{R})$ Wilson line. Second, it would be interesting to understand the physical interpretation of the two possible dressings of the bulk operator in both the metric and the Chern-Simons formulation of $3d$ gravity. Finally,  it would be nice to find a relation between our dressings, both of which are linear in the stress tensor, and the ``quadratic'' dressing recently put forth in \cite{Lewkowycz:2016ukf}. 

 It would also be interesting to check the relation between the bulk field \eqref{hkll} and the boundary OPE for the case of  non-universal contributions to the OPE, which do not follow by symmetry arguments alone. If the relation does continue to hold to all orders in $1/N$, and given that the geodesic integral can be viewed as a Mellin transform \cite{daCunha:2016crm}, then the  bulk field at an arbitrary distance inside the bulk would indeed be given by the (appropriately defined) inverse Mellin transform of the corresponding boundary OPE. It would be interesting to explore whether this leads to a new way to understand of the emergence of the holographic direction in AdS/CFT.

\bigskip

\noindent \textbf{Acknowledgements}

\vskip1mm
\noindent The author is grateful to Eliot Hijano, Nabil Iqbal, Elias Kiritsis, Per Kraus, Aitor Lewkowycz, Giuseppe Policastro and especially Dan Jafferis for useful conversations and correspondence. She would especially like to thank  Tom Hartman and Herman Verlinde for interesting discussions and collaboration on related topics.   Her work is supported by the  ERC Starting Grant 679278 Emergent-BH, the the Knut and Alice Wallenberg Foundation
under grant 113410212 (as Wallenberg Academy Fellow) and the Swedish Research Council  grant number 113410213.

\appendix

\section*{A. Bulk field in ``circular'' gauge}

It is of interest to obtain an explicit expression for the bulk field in a gauge different from radial gauge. In this appendix, we would like to obtain the dressing for a bulk operator connected via a Wilson line that stretches along a geodesic obeying
\be
z^2 + x^+ x^- = R^2\;, \;\;\;\; \frac{x^+}{x^-}= \a^2 \label{geodeq}
\ee
Letting $\tau^\mu$ be the tangent vector to this geodesic, we need to find the diffeomorphism $\xi_c$ that  asymptotes to \eqref{xiasy} at infinity and obeys $(\mathcal{L}_{\xi_c} g_{\mu\nu}) \tau^\nu =0$. While we need this condition to be satisfied only along some particular geodesic, we will request that it holds for all geodesics\footnote{These geodesics are semicircles in the Minkovski space conformal to AdS, hence the name for this gauge. Note that $g_{\tau \mu} =0$ is not a consistent gauge choice, as there exist physical configurations that cannot be brought into it. } that are symmetric around the origin $x^\pm=0$ and obey \eqref{geodeq} for some $R,\a$. 

Next, it is convenient to transform \eqref{poinc} to black string coordinates
\be
x^\pm = \sqrt{\frac{r^2-r_+^2}{r^2}} e^{2\pi T (\phi \mp t)} \;, \;\;\;\; z = \frac{r_+}{r} e^{2\pi T \phi}
\ee
in which the geodesic lines \eqref{geodeq} simply become lines of constant $t,\phi$. Their tangent vector is
\be
\tau= r \p_r = \frac{z^2}{x^+ x^-} (x^+ \p_+ + x^- \p_-) - z \p_z
\ee
The most general diffeomorphism that obeys the boundary conditions \eqref{xiasy} (we take $f_-(x^-) =0$ for simplicity) is
\bea
\xi_c & =& \left(f(\hat x^+ ) - x^+ \rho \, f'(\hat x^+ ) + \half  (x^+ \rho)^2 f''(\hat x^+ ) \right) \p_+ \non \\
&& \hspace{-1cm } - \, \frac{z^2}{2} f''(\hat x^+)\, \p_- + \frac{z}{2} (f'(\hat x^+) - x^+ \rho \, f''(\hat x^+)) \, \p_z
\eea
where we defined
\be
\hat x^+ = x^+ \sqrt{1+ \frac{z^2}{x^+ x^-}} \;, \;\;\;\;\; \rho=\sqrt{1+ \frac{z^2}{x^+ x^-}}-1
\ee
Note that $\hat x^+$ is nothing but the boundary endpoint of the geodesic that passes through the chosen bulk point, and  that $x^+ \rho = \hat x^+ - x^+$. Shifting the bulk field \eqref{fof} by  $\chi^\mu = \xi^\mu - \xi^\mu_c$,  the expression  we obtain for the linearized bulk field in this gauge is the same as \eqref{phihrad}, just with $x^+$ replaced with $\hat x^+$, in accordance with the fact that the endpoint of the Wilson line on the boundary has moved. It seems reasonable to assume that the expression will exponentiate, as before.


\begin{thebibliography}{99}


  \bibitem{Czech:2016xec} 
  B.~Czech, L.~Lamprou, S.~McCandlish, B.~Mosk and J.~Sully,
  ``A Stereoscopic Look into the Bulk,''
  JHEP {\bf 1607}, 129 (2016)
  arXiv:1604.03110 [hep-th].  
  
\vspace{-2mm}  
  
  \bibitem{deBoer:2016pqk}
  J.~de Boer, F.~M.~Haehl, M.~P.~Heller and R.~C.~Myers,
  ``Entanglement, holography and causal diamonds,''
  JHEP {\bf 1608} (2016) 162,
  arXiv:1606.03307 [hep-th].
  

\vspace{-2mm}   
  
\bibitem{daCunha:2016crm} 
  B.~Carneiro da Cunha and M.~Guica,
  ``Exploring the BTZ bulk with boundary conformal blocks,''
  arXiv:1604.07383 [hep-th].  
  
\vspace{-2mm} 

  \bibitem{Ferrara:1971vh} 
  S.~Ferrara, A.~F.~Grillo and R.~Gatto,
  ``Manifestly conformal covariant operator-product expansion,''
  Lett.\ Nuovo Cim.\  {\bf 2S2}, 1363 (1971)
  
\vspace{-2mm}   
  
\bibitem{Bena:1999jv}
  I.~Bena,
  ``On the construction of local fields in the bulk of AdS(5) and other spaces,''
  Phys.\ Rev.\ D {\bf 62} (2000) 066007,
  hep-th/9905186.  
  
\vspace{-2mm} 

\bibitem{Hamilton:2006az}
  A.~Hamilton, D.~N.~Kabat, G.~Lifschytz and D.~A.~Lowe,
  ``Holographic representation of local bulk operators,''
  Phys.\ Rev.\ D {\bf 74}, 066009 (2006),
  hep-th/0606141.    
  
  
\vspace{-2mm}  
  
\bibitem{Hamilton:2006fh} 
  A.~Hamilton, D.~N.~Kabat, G.~Lifschytz and D.~A.~Lowe,
  ``Local bulk operators in AdS/CFT: A Holographic description of the black hole interior,''
  Phys.\ Rev.\ D {\bf 75}, 106001 (2007),
   hep-th/0612053.  

\vspace{-2mm}   
  
\bibitem{Kabat:2011rz} 
  D.~Kabat, G.~Lifschytz and D.~A.~Lowe,
  ``Constructing local bulk observables in interacting AdS/CFT,''
  Phys.\ Rev.\ D {\bf 83}, 106009 (2011)
  arXiv:1102.2910 [hep-th].  
  

\vspace{-2mm}   
  
  \bibitem{Kabat:2016zzr}
  D.~Kabat and G.~Lifschytz,
  ``Locality, bulk equations of motion and the conformal bootstrap,''
  JHEP {\bf 1610} (2016) 091
  arXiv:1603.06800 [hep-th].
  

\vspace{-2mm}   
  
  \bibitem{Hijano:2015qja}
  E.~Hijano, P.~Kraus, E.~Perlmutter and R.~Snively,
  ``Semiclassical Virasoro blocks from AdS$_{3}$ gravity,''
  JHEP {\bf 1512} (2015) 077
  arXiv:1508.04987 [hep-th].


\vspace{-2mm}   

\bibitem{lamprou}  
L.~Lamprou and S.~McCandlish, to appear. See also L. Lamprou's talk at Perimeter
http://pirsa.org/displayFlash.php?id=16100060

\vspace{-1mm}   
  
  
  \bibitem{wdan}
  M.~Guica and D.~L.~Jafferis,
  ``On the construction of charged operators inside an eternal black hole,''
  arXiv:1511.05627 [hep-th].
  

\vspace{-2mm}   


\bibitem{Ferrara:1972uq} 
  S.~Ferrara, A.~F.~Grillo, G.~Parisi and R.~Gatto,
  ``The shadow operator formalism for conformal algebra. vacuum expectation values and operator products,''
  Lett.\ Nuovo Cim.\  {\bf 4}, 115 (1972)

\vspace{-2mm} 

%

  \bibitem{Heemskerk:2012mn} 
  I.~Heemskerk, D.~Marolf, J.~Polchinski and J.~Sully,
  ``Bulk and Transhorizon Measurements in AdS/CFT,''
  JHEP {\bf 1210}, 165 (2012)
  arXiv:1201.3664 [hep-th].
%
%
%
%
  
\vspace{-2mm} 
  \bibitem{Donnelly:2015hta} 
  W.~Donnelly and S.~B.~Giddings,
  ``Diffeomorphism - invariant observables and their nonlocal algebra,''
  Phys.\ Rev.\ D {\bf 93}, no. 2, 024030 (2016)
  arXiv:1507.07921 [hep-th].

\vspace{-2mm}   
 
 \bibitem{Skenderis:1999nb} 
 K.~Skenderis and S.~N.~Solodukhin,
  ``Quantum effective action from the AdS/CFT correspondence,''
  Phys.\ Lett.\ B {\bf 472}, 316 (2000),
    hep-th/9910023. 


%


\vspace{-2mm}   
  
\bibitem{Kabat:2013wga}
  D.~Kabat and G.~Lifschytz,
  ``Decoding the hologram: Scalar fields interacting with gravity,''
  Phys.\ Rev.\ D {\bf 89} (2014) no.6,  066010
  arXiv:1311.3020 [hep-th].  
  

\vspace{-2mm} 

\bibitem{Turiaci:2016cvo} 
  G.~Turiaci and H.~Verlinde,
  ``On CFT and Quantum Chaos,''
  JHEP {\bf 1612}, 110 (2016), 
  arXiv:1603.03020 [hep-th].

%
  
  

  


\vspace{-2mm} 

\bibitem{Fitzpatrick:2016mtp}
  A.~L.~Fitzpatrick, J.~Kaplan, D.~Li and J.~Wang,
  ``Exact Virasoro Blocks from Wilson Lines and Background-Independent Operators,''
  arXiv:1612.06385 [hep-th].

\vspace{-2mm} 

\bibitem{Fitzpatrick:2015foa} 
 A.~L.~Fitzpatrick, J.~Kaplan, M.~T.~Walters and J.~Wang,
  \emph{``Hawking from Catalan,''}, JHEP {\bf 1605}, 069 (2016), 
  arXiv: 1510.00014 [hep-th].


\vspace{-2mm}   
  
  \bibitem{Lewkowycz:2016ukf}
  A.~Lewkowycz, G.~J.~Turiaci and H.~Verlinde,
  ``A CFT Perspective on Gravitational Dressing and Bulk Locality,''
  arXiv:1608.08977 [hep-th].

  
\end{thebibliography}
\end{document}